\documentclass[lettersize,journal]{IEEEtran}
\usepackage{amsmath,amsfonts}
\usepackage{algorithmic}
\usepackage{algorithm}
\usepackage{array}
\usepackage[caption=false,font=normalsize,labelfont=sf,textfont=sf]{subfig}
\usepackage{textcomp}
\usepackage{stfloats}
\usepackage{url}
\usepackage{hyperref}
\usepackage{verbatim}
\usepackage{graphicx}
\usepackage{cite}
\usepackage{multirow}

\begin{document}

\title{Featured Trajectory Generation Module for TrackPuzzle}

	\author{Wanting Li~\IEEEmembership{Student Member,~IEEE,}, Yongcai Wang~\IEEEmembership{Member,~IEEE,}  Xiongfei Geng, Deying Li~\IEEEmembership{Member,~IEEE}
	% <-this % stops a space
	\thanks{Corresponding: Yongcai Wang, ycw@ruc.edu.cn}% <-this % stops a space
	\thanks{Manuscript received Janarary 8, 2022; revised xxx xx, 2022.}
}

\markboth{IEEE Transactions on Intelligent Transportation Systems,~Vol.~xx, No.~x, xxx~2022}%
{Li \MakeLowercase{\textit{et al.}}: Featured Trajectory Generation Module for TrackPuzzle}

\maketitle

\begin{abstract}
 Indoor route graph learning is critically important for autonomous indoor navigation. A key problem for crowd-sourcing indoor route graph learning is featured trajectory generation. In this paper, a system is provided to generate featured trajectories by crowd-sourcing smartphone data. Firstly, we propose a more accurate PDR algorithm for
 the generation of trajectory motion data. This algorithm uses ADAPTIV as the step counting method and uses the step estimation algorithm o make the trajectory more accurate
 in length. Next, the barometer is used to segment the tracks of different floors, and the track floors are obtained by WiFi feature clustering. Finally, by finding the turning point as the feature point of the trajectory, the vertices and edges of the trajectory are
 extracted to reduce the noise of the long straight trajectory.
\end{abstract}


\section{Introduction}

Global Positioning System (GPS) provides satisfied outdoor positioning services but does not work indoor. WiFi is currently the dominant local wireless network standard for
short-range communication and is the leading technology for indoor positioning\cite{li2017passively}. WiFi fingerprinting based on Received Signal Strength Indicator (RSSI) has been widely used in commercial indoor positioning systems because of its high positioning accuracy and ubiquitousness, such as Google and Apple. A traditional fingerprinting system requires a radio map comprised of a large quantity of WiFi fingerprints and each WiFi fingerprint contains a WiFi RSSI list and its ground truth location label. To build up radio maps, professionals consume huge amount of time to conduct a site survey for the ground truth location labels\cite{li2015time}. To avoid the intensive training efforts in fingerprinting, crowdsourcing approaches have become increasingly interesting for researchers. And then, we use crowdsourcing smartphone data to build indoor map. 

In this work, we propose a novel system for automatic construction of featured crowdsourcing trajectory generation, which requires neither priori knowledge of a floor
plan nor manual initializations of PDR traces. Our contributions are summarized as follows.

Firstly, daptive jerk pace step detection algorithm is exploited for step detection; hierarchical support vector machine method is exploited for displacement estimation; and magnetometer aided heading estimation is exploited in PDR. They work together to infer the relative motions more accurately from the raw IMU data sequences. 

Then, a floor classification algorithm based on jointly DBSCAN clustering of barometer readings of PDR trajectories and WiFi MAC address clustering is proposed. It can divide the trajectories into correct floors rather reliably. 

Finally, eliable turning point detection method is proposed to detect reliable turnings as the motion feature points. Small turnings are filtered out and frequent turning segments are deleted to exclude the impacts of irregular motions. So each PDR trajectory is converted to a \emph{chain type graph}, which uses the end points and the extracted turning points as vertices and uses the motion vectors between vertices as edge.

\section{System Overview}

Trackpuzzle design is based on a participatory sensing approach, where phone sensors measurements collected from users moving naturally inside the building are processed to generate featured motion traces which in turn can provide an adequate description of the indoor route graph. In this paper, we only provide the method of featured trajectory generation, and the part of indoor route graph generation is discussed in another paper. The system consists of two main module: (a) the Data Collection Module is responsible for collecting measurements from users’ devices, and (b) the Featured Trajectory Generation Module is responsible for building accurate motion traces based on Motion Trajectory Generation, Floor Classification and Vertex \& Edge Extraction.  

\begin{figure}
	\includegraphics[width=1\linewidth]{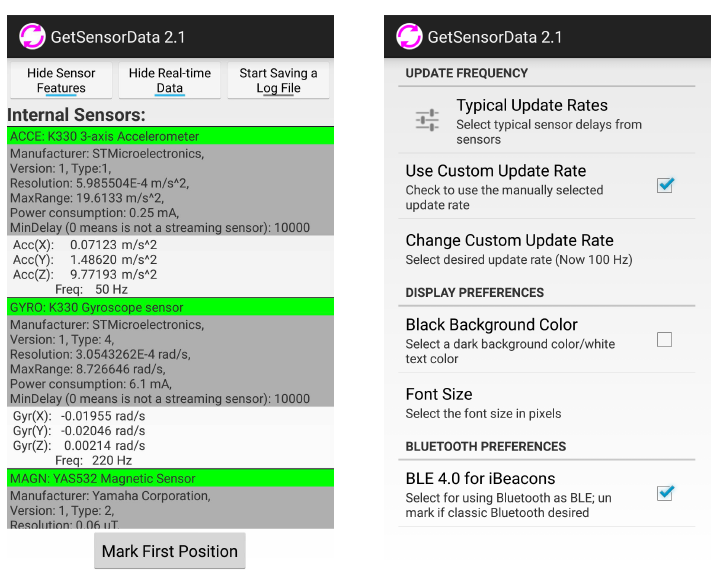}
	\caption{Getsensordata APP can be used to collect multi-sensors data from smartphone.}
	\label{fig:getsensor}
\end{figure}

The data collection module collects measurements from phone embedded sensors including: accelerometer, magnetometer, gyroscope, barometer and the received cellular and WiFi signal strengths from the available cell towers and WiFi access points. In this system, we use Getsensordata app\cite{jimenez2019tools} to collect these data as shown in Figure \ref{fig:getsensor}, and then generate files and send them back to the server for processing.

\section{Featured Trajectory Generation Module}

In this section, we will firstly introduce (1) how a PDR algorithm is carefully designed to recover the motion trajectories as accurate as possible; (2) how the motion trajectories are segmented into different floors; and (3)  how the motion trajectories are filtered and processed to extract reliable turning points as the feature points, so that each motion trajectory is converted to a chain-type graph using the feature points as vertices and motion vectors between feature points as edges. 

\subsection{Motion Trajectory Generation}
% LI wanting
The first step is to recover the motion trajectory from the raw IMU data, which relies on a dead-reckoning based approach. The user's current location $(X_k,Y_k)$ is estimated from the previous location $(X_{k-1},Y_{k-1})$, the estimated step length is denoted by $S$, and the estimated motion direction is denoted by $\theta $.  Then: 
\begin{equation}
\begin{gathered}
{{\text{X}}_{\text{k}}}{\text{  =  }}{{\text{X}}_{{\text{k - 1}}}}{\text{  +  S*cos(}}\theta {\text{)}} \hfill \\
{{\text{Y}}_{\text{k}}}{\text{  =  }}{{\text{Y}}_{{\text{k - 1}}}}{\text{  +  S*sin(}}\theta {\text{)}} \hfill \\ 
\end{gathered} 
\end{equation}
where the displacement $S$ is mainly obtained from the accelerometer data, while $\theta$ is estimated from the magnetometer and the gyroscope data.  The PDR algorithm has been extensively studied in the literature. We conduct extensive experiments and select the following methods to try to recover the motion trajectories accurately. 

%Theoretically, the user's travelled distance can be calculated by integrating the acceleration and time twice. However, due to the existence of accelerometer noise and the earth gravity component in the acceleration signal, the error will accumulate rapidly over time. The displacement error increases cubic with time and can reach 100 meters after one minute operation. 
\subsubsection{Step Detection}
The accuracy of step detection  affects  the accuracy of PDR greatly.  Step detection is based on the detection of  peak and valley oscillation on the magnitude signal of the  accelerometer data. 
Because the motion trajectories may be collected by phones using different brands of sensors and the users may take the phones in different places and poses, the magnitudes of the  accelerometer  data are different in different trajectories, or even different in different segments of one trajectory. 

To detect steps correctly in different application scenarios, we uses ADAPTIV\cite{murray2018adaptiv}, an adaptive jerk pace buffer based step detection algorithm. 
Jerk is the acceleration change from peak to valley  in a step oscillation.  Pace is the time duration of a step. ADAPTIV proposes adaptive jerk and adaptive pace thresholds,   and proposes a buffer to update the jerk and pace thresholds.  
It shows reliable step detection accuracy for accelerometer data collected by different phone poses, which provides better accuracy than the step detection algorithm implemented in Android system.  

\begin{figure}[t]	
		\includegraphics[width=1\linewidth]{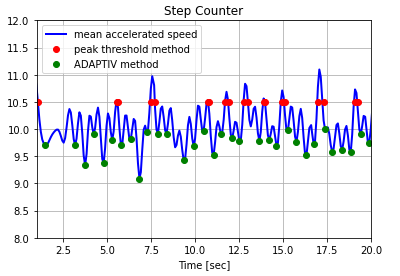}
		\caption{Red points are results by peak threshold method which miss many steps. Green points are results by ADAPTIV method, the jerk threshold is dynamically updated.}
		\label{fig:jerk}
	%\caption{text}
\end{figure}

%	The method is designed to make the step detection threshold be adaptive to the user's gait.  We accomplish this by dynamically updating a step jerk average threshold, each time a step is detected. 
%	\color{blue}
\subsubsection{Stride Length Estimation}
Stride length estimation is to estimate the displacement $S$ in each step.  \cite{alzantot2012uptime}  shows that the stride
length changes significantly for the same person when the
gait type changes.  
Therefore, this paper exploits the  variable stride length estimation method proposed in  UPTIME \cite{alzantot2012uptime} to improve the accuracy of stride length estimation. 

The method uses a hierarchical support vector machine\cite{alzantot2012uptime} to design a gait classifier. 
The signal characteristics in a step are extracted to construct features, including the stride duration, accelerometer signal variance, acceleration peak, RMS acceleration etc.  
Then the user gait is classified by the hierarchical support vector machine using these features as input,  and the stride length is estimated by looking up table using the estimated gait.  
The method can much improve the accuracy and adaptivity of stride length estimation. 
\subsubsection{Direction Estimation}
Direction estimation is to estimate the heading angle $\theta$. 
We exploit HUMAINE \cite{mohssen2014s},  a sensor fusion method to accurately  estimate user orientation relative to the Earth coordinate frame. 

In HUMAINE, firstly, the gravity vector is tracked by  the opportunities when the acceleration vector approximately equals $g$(9.80665m/s2) and by fusion of gyroscope and magnetometer data.  Whenever the trend of gyroscope and magnetometer reading is correlated, the magnetometer reading is believed accurate\cite{zhou2014use} and is selected to estimate  the phone's direction.  

By tracking gravity vector and the correct magnetometer readings,  equations for estimating roll and pitch are set up by relating phone pose with the gravity vector. Then the equation to estimate yaw angle (phone's orientation) is set up based on the magnetometer readings.    
But the phone's orientation maybe different from the user's motion orientation. So a principal component analysis (PCA) method is applied on the extracted  linear acceleration data of the phone to find the direction with the largest variance. The first principle direction with acute angle with the phone orientation is then detected as the user moving orientation. 

%forced out from  the acceleration to obtain the linear acceleration component in the phone's coordinate system. Intuition is that the trend of magnetometer and gyroscope is the correct reading of magnetometer; If the magnetometer is not reflecting the gyroscope trend, it may be affected by electronic interference. Therefore, we apply principal component analysis (PCA) to apply linear acceleration, and use the direction of the first principal component (with maximum variance) as the user direction.

\subsection{Floor Classification}
% LI wanting
% add example graphs. 
The first step to register the PDR trajectories is to find the trajectories that are collected from the same floor.  One trajectory maybe collected when a user walks through multiple floors. Such trajectories should be cut into  segments of different floors for registering  the trajectory segments of the same floor to recover the floor routes. 

%The data we collect may come from different areas and floors of the same building, and there may be only one or several walking edges between these areas. In fact, the number of trajectories we walk in the same area is much larger than that of different regions. From the HGO method, we can see that we can calculate the tight sub graph separately and finally adjust the final 3D map. 
The barometer reading is an important clue to distinguish floor index. However, barometer sensors have good accuracy in measuring relative pressure changes, but have unsatisfied accuracy in measuring absolute pressure. This is because the biases of barometer readings are different in different phones.  We therefore can use barometer readings to confidently separate a trajectory into segments of different floors but the barometer readings cannot tell exactly the floor index of the segments. 

We find that the trajectories on the same floor share more common MAC addresses of the WiFi APs, and the trajectories on different floors share much less common MAC addresses. We therefore propose to use the Jaccard coefficient of MAC address appearing in trajectory segments to further classify the trajectory segments to classify their floor indices.

%We found that we can easily divide the trajectories across multiple layers into different regions by DBSCAN algorithm. But because of the influence of humidity and temperature on the barometer results, there is no way to cluster the segmented trajectories, that is to say, the trajectories correspond to a certain floor. 
% add some details. 

Firstly, the PDR points in a trajectory is clustered by DBSCAN method regarding the barometer readings of the points, which automatically cluster the trajectory data into multiple clusters if the trajectory travels multiple floors.  Each cluster is a trajectory segment in one floor. %But we cannot tell exactly the floor index of the trajectory segment. 

\begin{figure*}[t]
	\begin{minipage}[t]{0.48\linewidth}
		\centering
		\includegraphics[width=2.5in]{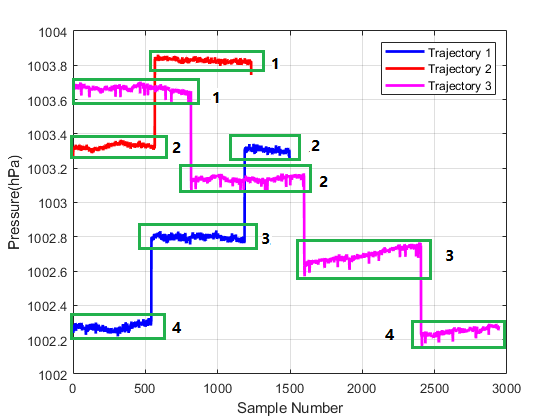}
		\caption{Floors classification using barometer and WIFI signal values.}
		\label{fig:floor}
	\end{minipage}%
	\hspace{0.1in}
	\begin{minipage}[t]{0.48\linewidth}
		\centering
		\includegraphics[width=2.4in]{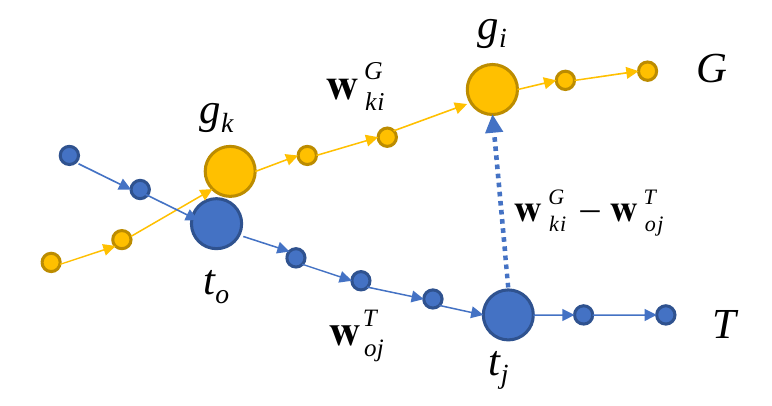}
		\caption{Calculate the spatial difference between $g_i$ and $t_j$ based on the associated preceding nodes .}
		\label{figtan}
	\end{minipage}
\end{figure*}

Then all the trajectory segments separated from different trajectories  are put together. For each trajectory segment (say trajectory segment $s_i$), the set of AP MAC address appearing in $s_i$ is extracted and is denoted by $f_i$. Then for any two trajectory segments $s_i$ and $s_j$, we have the following two MAC address sets: 
\begin{equation}
\begin{gathered}
{f_i} = \{ ma{c_{i1}},ma{c_{i2}},...,ma{c_{im}}\}  \hfill \\
{f_j} = \{ ma{c_{j1}},ma{c_{j2}},...,ma{c_{jn}}\}  \hfill \\ 
\end{gathered} 
\end{equation}
where $mac$ is the MAC address of an AP. We then obtain the intersection and union of  $f_i$ and $f_j$ .
\begin{equation}
\begin{gathered}
MA{C_{\operatorname{int} }} = {f_i} \cap {f_j} \hfill \\
MA{C_{uni}} = {f_i} \cup {f_j} \hfill \\ 
\end{gathered} 
\end{equation}
The Jaccard coefficient of $f_i$ and $f_j$ is calculated using the following equation:
\begin{equation}
Ja{c_{ij}} = |MA{C_{\operatorname{int} }}|/|MA{C_{uni}}|
\end{equation}
where $|X|$ is the carnality of the set $X$. 
Finally, using jaccard coefficient $Ja{c_{ij}}$ as a distance, a 
hierarchical clustering\cite{johnson1967hierarchical} is performed on the trajectory segments to obtain specific floor indices. The final cluster of trajectory segments with the highest average pressure is the trajectories of the first floor. An example is shown in Fig.\ref{fig:floor}.  The three trajectories are segmented and are clustered into trajectory segments of four floors. 

%	The merging algorithm of hierarchical clustering calculates the similarity between two kinds of data points, combines the most similar data points among all data points, and iterates the process repeatedly. In short, the merging algorithm of hierarchical clustering is to calculate the distance between each class of data points and all data points to determine the similarity between them. The smaller the distance is, the higher the similarity is. The nearest two data points or categories are combined to generate a clustering tree. Here we use 
%	

\subsection{Vertex and Edge Extraction On Each Trajectory}
Then we consider feature extraction on each trajectory segment. 
Traditional RSS features are noisy.  But different PDR trajectories on a same floor show rather stable relative turning patterns  at common turning points. We therefore propose to extract the reliable \emph{turning points (TPs)}  on the trajectories. The extracted turning points become vertices. The RSS at the turning point is used as vertex feature and the motion vectors among the turning points will become edges. This will turn each trajectory segment into a chain-type graph.  

\subsubsection{Turning Points Detection} 

\begin{figure*}[!h]	
		\begin{minipage}[t]{0.33\linewidth}
			\centering
			\includegraphics[width=2.1in]{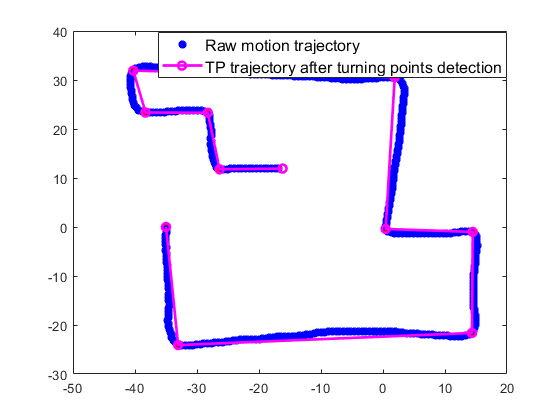}
			\caption{Normal-An example of turning point detection.}   
			\label{figline2}
		\end{minipage}%
		\begin{minipage}[t]{0.33\linewidth}
			\centering
			\includegraphics[width=2.1in]{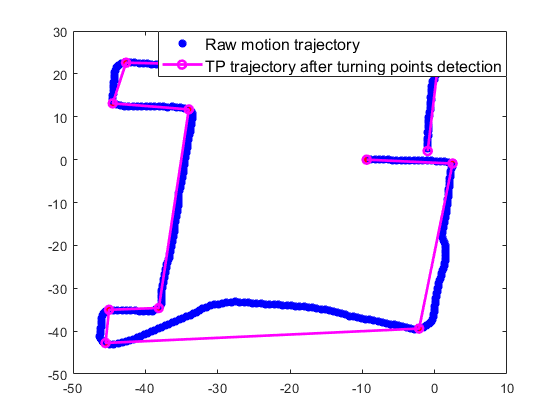} 
			\caption{Slowly and noisy turning-Tolerate turning noises.}  
			\label{figline4}
		\end{minipage}%
		\begin{minipage}[t]{0.33\linewidth}
			\centering
			\includegraphics[width=2.1in]{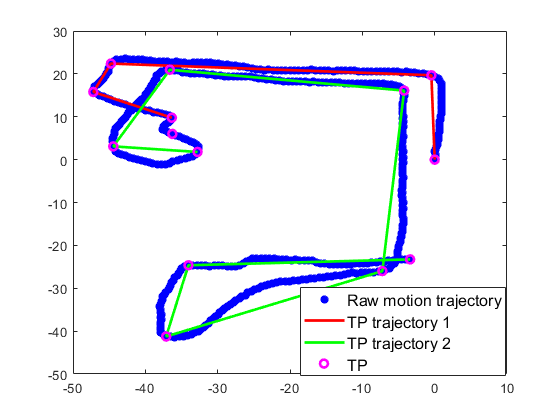}
			\caption{Noisy PDR with frequent turnings-Decomposed into two sub-trajectories.}   
			\label{figline5}
		\end{minipage}%
	       
\end{figure*}

\begin{figure*}[!h]	
	\begin{minipage}[t]{0.5\linewidth}
		\centering
		\includegraphics[width=2.5in]{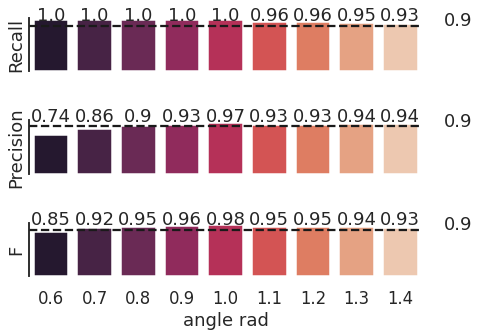}   
		\label{fig:turnangle}
	\end{minipage}%
	\begin{minipage}[t]{0.5\linewidth}
		\centering
		\includegraphics[width=2.5in]{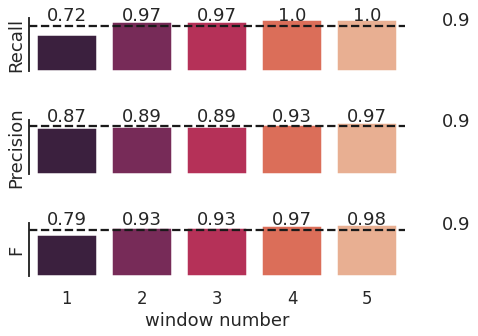}
		\label{fig:turnwindow}
	\end{minipage}%
	\caption{Parameter selection for turning point detection.(a)Recall,precision and F of different angle rads.(b)Recall,precision and F of different window numbers.}
	\label{fig:turn}
\end{figure*}
%Refer to \textit{Harris Corner Detection},  when walking along the edge,  $\theta$ will change a lot at the corner,  but not much at other times. So turnings with sharp change in angle are regarded as POIs. Our problem is to seek out the POIs by PDR sequences. 
The ideal turning point detection is shown in Fig.\ref{figline2}, where reliable turning points are detected as feature points.  Because user motions and the sensor data have noises, there are heading direction noises in each step, as shown in Fig.\ref{figline4}.  To detect reliable turning points, we need on one hand to avoid  false alarming caused by heading noises, and on the other hand to reliably detect turning when the accumulated heading direction has changed enough.  

We therefore design a sliding window based turning point detection algorithm.  The algorithm starts by using the first point as the origin. $\bf V$ is the detected set of turning points which is initialized by $\bf V={1}$.  
Starting from $(x_1, y_1)^T$, whenever a turning point (denoted as point $i$) is detected, this turning point will become the starting point of a new window, and is set as the origin of the new window. For each point following this point, i.e.,  ${i+1, \cdots, i+t}$, turning angle $\{\alpha_{i+1}, \cdots, \alpha_{i+t}\}$ are calculated by
\begin{equation}
{\alpha _{i + t}} = \left| {\operatorname{atan} \left( {\frac{{{y_{i + t}}}}{{{x_{i + t}}}}} \right) - \operatorname{atan} \left( {\frac{{{y_i}}}{{{x_i}}}} \right)} \right|, \textrm{   } t=1, 2, \cdots
\end{equation} 
When ${\alpha _{i + t}} \le \varepsilon$, $i+t$ is added to the current window. Only when ${\alpha _{i + t}} >\varepsilon$, $i+t-1$ will be detected as a new turning point. $i+t-1$ is added into the turning point set. The window length is reset to 1 and the point $i+t-1$ becomes the new origin. 
By this method, the sliding window filters out the turning noises and detects a turning point when enough  turning has been accumulated.  For a trajectory segment $\mathbf T_i$, all detected turning points will be extracted as vertices, denoted by set $\mathbf V_i$. The motion vector between each pair of vertices is a motion edge. The edges form a edge set $\mathbf E_i$.  So the trajectory  $\mathbf T_i$ is represented by a chain-type graph $\mathbf G_i=(\mathbf V_i, \mathbf E_i)$. 
%The detailed algorithm is given in  Algorithm~\ref{algorithm1}. 

\subsection{Delete Frequent Turning Segments } 
Turning points will be detected frequently if a user turns frequently during data collection. 
Such kinds of trajectories cannot  accurately reveal the indoor path features, because the turnings are mainly caused by the user's irregular motions.  Since there are redundant trajectories, we don't need these ``bad" trajectories.  
We therefore propose to delete trajectory segments with continuous  turning points. If some turning points appear very closely on a trajectory, such sub-trajectories will be deleted to avoid the frequent turnings' impacts to the trajectory registration.  We propose to: 
\begin{enumerate}
	\item Divide a trajectory into two trajectories if two turning points are detected successively. 
	\item Delete a sub-trajectory if the length of a sub-trajectory is too short. 
\end{enumerate} 
The algorithm
% is given in Algorithm~\ref{algorithm2} .  The input of the algorithm is a trajectory graph $\bf G=(V, E, W, Z)$ with extracted vertices and edges. It 
checks the vertex set  sequentially. If two neighboring vertices $v_i, v_{i+1}$ are not successive in the original trajectory, the vertices are added into current trajectory graph $\mathbf{G}_i$. If two neighboring vertices are successive, a new sub-trajectory graph $\mathbf{G}_{i+1}$ will be generated and using $v_{t+1}$ as the starting point of this new sub-trajectory graph.  If $\mathbf G_i$ is too short, $\mathbf G_i$ will be deleted, since the short sub-trajectories can not well reveal the path structure. %The detailed algorithm is given in Algorithm~\ref{algorithm2}. 
An example is shown in Fig.\ref{figline5}. 

\subsection{The Motion Vector and the RSS Vertex Feature}
After forming each chain-type graph $\mathbf G_i=(\mathbf V_i, \mathbf E_i)$, the motion vector between each pair of vertices is calculated. 
Note that $v_i$ stores the $i$th vertex's  origin index in the  trajectory. Let $\mathbf{w}_{j-1, j}$ be the motion vector between $v_{j-1}$ and $v_j$. Then

\begin{equation}
\label{equedge}
\mathbf{w}_{j-1, j} = {\vec u_{v_{j-1}}}  +{\vec u_{v_{j-1}+1}} + ... + {\vec u_{v_{j}-1}}
\end{equation}
where ${\vec u_{v_{j-1}}} $ is the motion vector in the original trajectory from node $v_{j-1}$ to node $v_{j-1}+1$.     
Note  the summation of motion vectors between successive turning points smooths the step-by-step motion noises. The motion vector gives the motion estimation between each pair of vertices. 
We then use $\mathbf Z_i=\{z_{v_1}, z_{v_2}, \cdots, z_{v_{n_i}}\}$ to represent the RSS signature of vertices.   
So a trajectory segment $\mathbf T_i$ is converted to a \emph{chain-type graph} $\bf G_i=(V_i, E_i, W_i, Z_i)$, 
%, where $\mathbf{V}_i = \{v_1, v_2, \cdots, v_{n_i}\}$. $\mathbf{E}=\{(v_i, v_{i+1})\}, i=1, \cdots, n_i-1$, $\mathbf{W}=\{\mathbf w_{1, 2}, \mathbf w_{2, 3}, \cdots, \mathbf w_{n_i-1, n_i} \}$.
where $\bf V_i$ are reliable turning points; $\bf W_i$ represent the motion vectors of edges; and  $\mathbf Z_i$ represents the RSS features of vertices.  Note that each trajectory graph consists only one path, which likes a chain.

\section{Experiment Results for Track Generation}

\subsection{Performance for Floor Classification}

First, we analyze the performance of DBSCAN algorithm for separating the trajectories to different floors. We set the maximum number of clusters in DBSCAN as 20.  %The outliers without successful clustering are the sampling points of up and down stairs. Then we divide a single trajectory, 
For each dataset, we put all trajectories together for floor clustering. By comparing with the ground truth, the floor classification accuracy is shown $100\%$ for all the trajectories in each data-set. The results are shown in Table~\ref{tab:floor}.

\begin{table}[h]  
	
	\centering  
	%\fontsize{8.5}{8}\selectfont  
	
	\caption{Floor Classification Accuracy}  
	\label{tab:floor}  
	\begin{tabular}{p{2.3cm}p{2cm}p{1.1cm}}  
		\hline  
		Data-set & Floor number&Accuracy\\   
		\hline  
		Dangdai Mall&3&$100\%$\\
%		Xinmate&1&$100\%$  \\
		Atlantis le Centre&3&$100\%$\\ 
		
		\hline 
	\end{tabular}  
\end{table}  
 
%\begin{table}[h]  
%	
%	\centering  
%	%\fontsize{8.5}{8}\selectfont  
%	
%	\caption{Floor Classification Accuracy}  
%	\label{tab:floor}  
%	\begin{tabular}{p{2.3cm}p{2cm}p{1.1cm}}  
%		\hline  
%		Data-set & Floor number&Accuracy\\   
%		\hline  
%		Dangdai Mall&3&$100\%$\\
%		Xinmate&1&$100\%$  \\
%		Atlantis le Centre&3&$100\%$\\ 
%		
%		\hline 
%	\end{tabular}  
%\end{table}  

\subsection{Optimizing the Parameters for Turning Point Detection}
The turning detection threshold ${\varepsilon }$ and sliding window length $t$ of the turning point detection algorithm are optimized using the experimental data. %Since the trajectories in data-sets contain the ground truth of the turning points, 	we adjust the parameters by optimizing the turning point detection accuracy. 
For evaluating turning point detection accuracy, we  use precision, recall and F-measure as the evaluation indices for turning point detection. 
%	Let's denote:
%	\begin{itemize}
%		\item \emph{TPE}: predict real TPs as TPs;
%		\item \emph{FNE}: predict real TPs as not TPs ;
%		\item \emph{FPE}: predict real not TPs  as TPs;
%		\item \emph{TNE}: predict real not TPs as not TPs. 
%	\end{itemize} 

%	So the \emph{precision}, \emph{recall} and \emph{F-measure} are defined as following: %\emph{Precision}: represents the classification effect (precision effect) of the classifier. It predicts the correct frequency value in the instance with TPs prediction where
%	$precision{=\frac{{TPE}}{{TPE + FPE}}}$, $recall{\text{ = }}\frac{{TPE}}{{TPE + FNE}}$. The \emph{F-measure} is used to measure precision and recall, which is the harmonic mean of the two values where $F{\text{ = }}\frac{{2precision*recall}}{{precision + recall}}$.

Using the dataset of Dangdai Shopping Mall, we evaluated the \emph{precision, recall} and \emph{F} of turning point detection when $\varepsilon$ varies is the range $[0.6,1.4]$. The results are shown in  Fig(\ref{fig:turn}(a)). It can be seen that ${\varepsilon =1 }$ obtains the highest $F-measure$. Then, by fixing  ${\varepsilon =1 }$, we calculated the \emph{precision, recall} and \emph{F} of as a function of the window length $t$ when $t$ varies in $[1,5]$. The result is shown in Fig(\ref{fig:turn}(b)). It can be seen that when  $t=4$, the \emph{precision} reaches 100\%.  So turning point detection parameters are set as $t=4, \varepsilon =1$ in the final experiments. 

\bibliographystyle{IEEEtran}
\bibliography{references.bib}

\begin{thebibliography}{1}
\providecommand{\url}[1]{#1}
\csname url@samestyle\endcsname
\providecommand{\newblock}{\relax}
\providecommand{\bibinfo}[2]{#2}
\providecommand{\BIBentrySTDinterwordspacing}{\spaceskip=0pt\relax}
\providecommand{\BIBentryALTinterwordstretchfactor}{4}
\providecommand{\BIBentryALTinterwordspacing}{\spaceskip=\fontdimen2\font plus
\BIBentryALTinterwordstretchfactor\fontdimen3\font minus
  \fontdimen4\font\relax}
\providecommand{\BIBforeignlanguage}[2]{{%
\expandafter\ifx\csname l@#1\endcsname\relax
\typeout{** WARNING: IEEEtran.bst: No hyphenation pattern has been}%
\typeout{** loaded for the language `#1'. Using the pattern for}%
\typeout{** the default language instead.}%
\else
\language=\csname l@#1\endcsname
\fi
#2}}
\providecommand{\BIBdecl}{\relax}
\BIBdecl

\bibitem{li2017passively}
Z.~Li and T.~Braun, ``Passively track wifi users with an enhanced particle
  filter using power-based ranging,'' \emph{IEEE Transactions on Wireless
  Communications}, vol.~16, no.~11, pp. 7305--7318, 2017.

\bibitem{li2015time}
Z.~Li, T.~Braun, and D.~C. Dimitrova, ``A time-based passive source
  localization system for narrow-band signal,'' in \emph{2015 IEEE
  International Conference on Communications (ICC)}.\hskip 1em plus 0.5em minus
  0.4em\relax IEEE, 2015, pp. 4599--4605.

\bibitem{jimenez2019tools}
A.~R. Jim{\'e}nez, F.~Seco, and J.~Torres-Sospedra, ``Tools for smartphone
  multi-sensor data registration and gt mapping for positioning applications,''
  in \emph{2019 International Conference on Indoor Positioning and Indoor
  Navigation (IPIN)}.\hskip 1em plus 0.5em minus 0.4em\relax IEEE, 2019, pp.
  1--8.

\bibitem{murray2018adaptiv}
D.~Murray and R.~Bonick, ``Adaptiv: An adaptive jerk pace buffer step detection
  algorithm,'' 2018.

\bibitem{alzantot2012uptime}
M.~Alzantot and M.~Youssef, ``Uptime: Ubiquitous pedestrian tracking using
  mobile phones,'' in \emph{2012 IEEE Wireless Communications and Networking
  Conference (WCNC)}.\hskip 1em plus 0.5em minus 0.4em\relax IEEE, 2012, pp.
  3204--3209.

\bibitem{mohssen2014s}
N.~Mohssen, R.~Momtaz, H.~Aly, and M.~Youssef, ``It's the human that matters:
  Accurate user orientation estimation for mobile computing applications,''
  \emph{arXiv preprint arXiv:1411.2156}, 2014.

\bibitem{zhou2014use}
P.~Zhou, M.~Li, and G.~Shen, ``Use it free: Instantly knowing your phone
  attitude,'' in \emph{Proceedings of the 20th annual international conference
  on Mobile computing and networking}, 2014, pp. 605--616.

\bibitem{johnson1967hierarchical}
S.~C. Johnson, ``Hierarchical clustering schemes,'' \emph{Psychometrika},
  vol.~32, no.~3, pp. 241--254, 1967.

\end{thebibliography}
\newpage

% your own custom macro containing the $\backslash${\tt{includegraphics}} command to make things
%\bf{If you include a photo:}\vspace{-33pt}

	\section{Biography Section}
%	If you have an EPS/PDF photo (graphicx package needed), extra braces are
%	needed around the contents of the optional argument to biography to prevent
%	the LaTeX parser from getting confused when it sees the complicated
%	$\backslash${\tt{includegraphics}} command within an optional argument. (You can create
%	your own custom macro containing the $\backslash${\tt{includegraphics}} command to make things
%	simpler here.)

\vspace{11pt}
\begin{IEEEbiography}
	[{\includegraphics[width=1in,height=1in,clip,keepaspectratio]{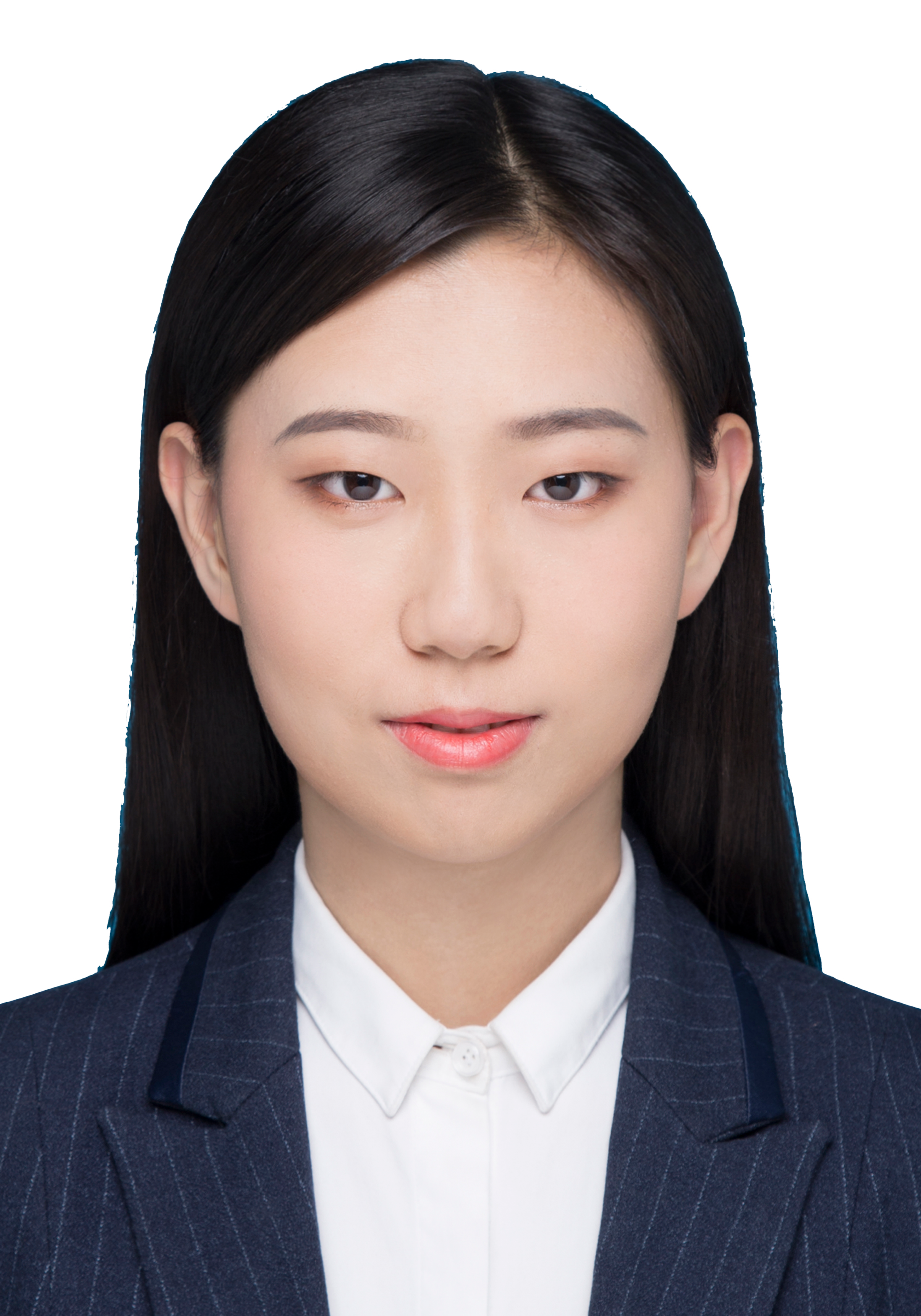}}]{Wanting Li}is a PhD in the Department of Computer Science, Renmin University of China.
	She received the BS degree in Computer science and technology from China Agricultural University (2019). She won Honorable Mentions in 2018 mathematical modeling contest of American College Students. Her research is unsupervised learning pervasive computing. 
\end{IEEEbiography}
\vskip -2\baselineskip plus -1fil
\begin{IEEEbiography}
	[{\includegraphics[width=1in,height=1in,clip,keepaspectratio]{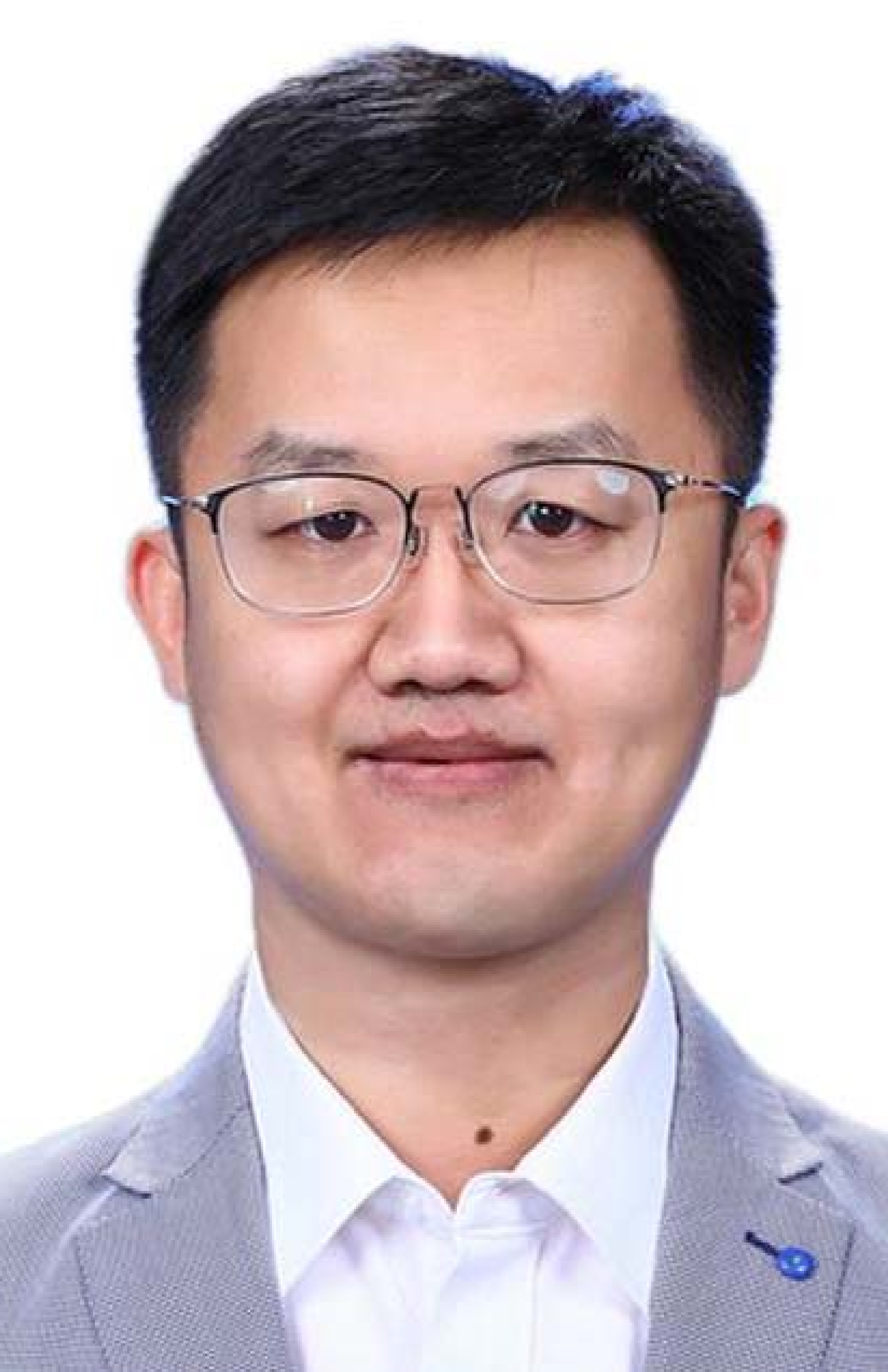}}]{Yongcai Wang}
	received his BS and PhD degrees from department of automation sciences and engineering, Tsinghua University in 2001 and 2006 respectively. He worked as associated researcher at NEC Labs. China from 2007-2009. He was an research scientist in Institute for Interdisciplinary Information Sciences (IIIS), Tsinghua University from 2009-2015. He was a visiting scholar at Cornell University in 2015. He is currently associate professor at  Department of Computer Sciences, Renmin University of China.  His research interests include network localization, and  combinatorial optimization and applications.   
\end{IEEEbiography}
\vskip -2\baselineskip plus -1fil
\begin{IEEEbiography}
	[{\includegraphics[width=1in,height=1in,clip,keepaspectratio]{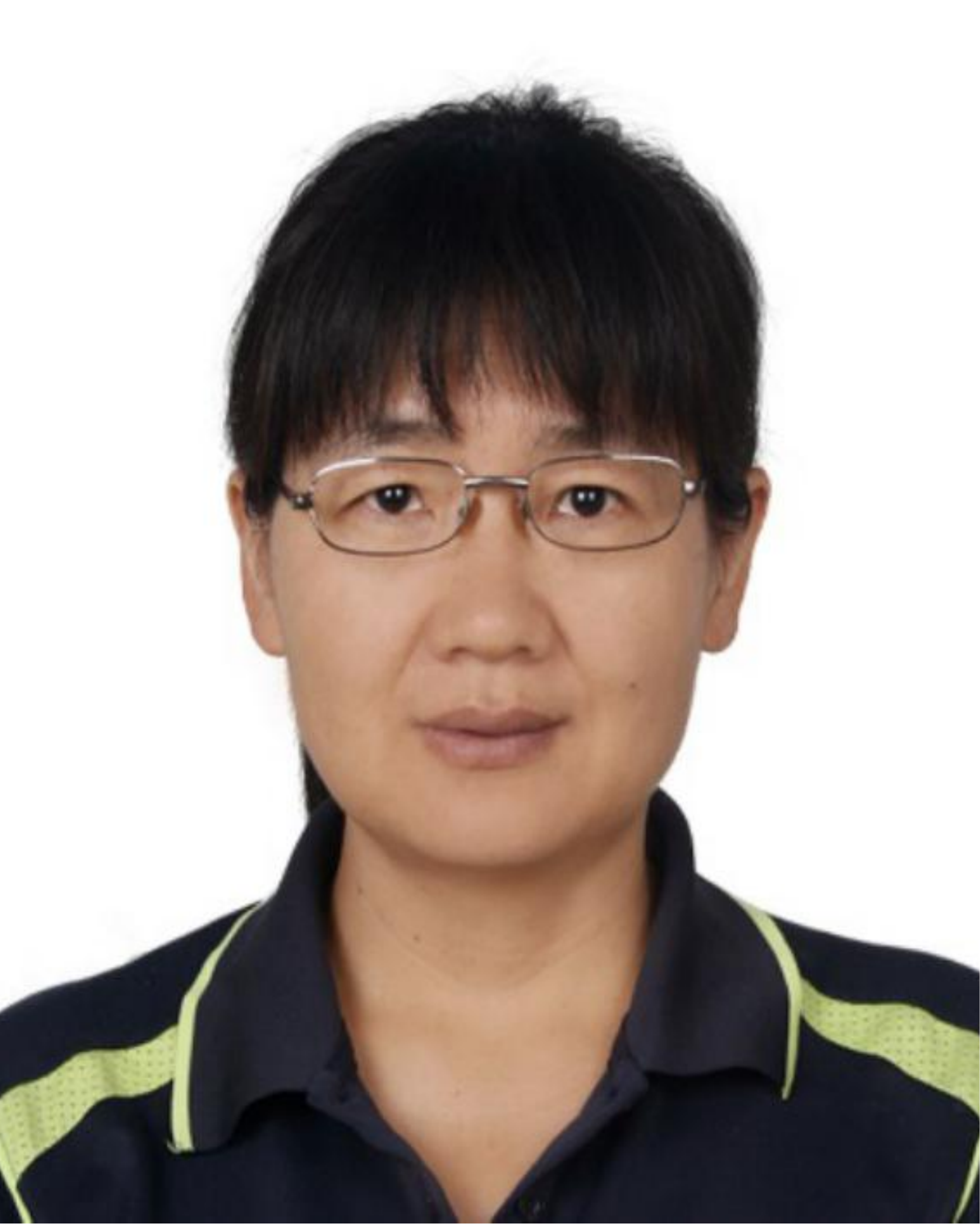}}]{Deying Li}
	received the MS degree in Mathematics from Huazhong Normal University (1988)
	and PhD degree in Computer Science from City University of Hong Kong (2004). She is 
	currently a Professor in the Department of Computer Science, Renmin University of China. Her
	research includes wireless networks, mobile computing, social network and algorithm design and analysis. 
\end{IEEEbiography}
\vskip -2\baselineskip plus -1fil

\vfill

\end{document}